# Merger-driven or Internal Evolution? A New Morphological Study of Tidal Disruption Event Host Galaxies

Janet N.Y. Chang[1,2], Connor Bottrell[3], Lixin Dai[1,2], Rudrani Kar Chowdhury[1,4], Meng Gu[2,5], Renbin Yan[6,7], Leonardo Ferreira[8,9], Sara L. Ellison[8], Scott Wilkinson[8], and Thomas de Boer[10]
[1] Department of Physics, The University of Hong Kong, Pokfulam Road, Hong Kong, People's Republic of China; lixindai@hku.hk
[2] The Hong Kong Institute for Astronomy and Astrophysics, The University of Hong Kong, Pokfulam Road, Hong Kong, People's Republic of China
[3] International Centre for Radio Astronomy Research, University of Western Australia, 35 Stirling Hwy, Crawley, WA 6009, Australia
[4] Tata Institute of Fundamental Research, Homi Bhabha Road, Mumbai 400005, India
[5] Department of Astronomy, Tsinghua University, Beijing 100084, People's Republic of China
[6] Department of Physics, The Chinese University of Hong Kong, Shatin, New Territories, Hong Kong, People's Republic of China
[7] CUHK Shenzhen Research Institute, No.10, 2nd Yuexing Road, Nanshan, Shenzhen 518000, People's Republic of China
[8] Department of Physics and Astronomy, University of Victoria, Victoria, BC V8P 5C2, Canada
[9] Centre for Astronomy and Particle Theory, University of Nottingham, Nottingham NG7 2RD, UK
[10] Institute for Astronomy, University of Hawaii, 2680 Woodlawn Drive, Honolulu, HI 96822, USA
*Received 2026 February 13; revised 2026 April 23; accepted 2026 May 12; published 2026 June 8*

## Abstract

The host galaxies of tidal disruption events (TDEs) show enhanced central stellar concentration and are preferentially found in poststarburst and green valley populations. This connection has led to the proposal that TDE host galaxies likely have gone through recent mergers. We conduct a new morphological study of 14 TDE host galaxies, using the $r$-band images from the Sloan Digital Sky Survey (SDSS), Dark Energy Camera Legacy Survey, and Ultraviolet Near-Infrared Optical Northern Survey, with the images from the latter two surveys having much higher depth and resolution than SDSS. We examine galaxy structures using conventional methods and also apply diagnostics of merger activity from a suite of machine learning models. Consistent with previous studies, our results show that TDE host galaxies are ∼16% more centrally concentrated when compared to non-TDE-host controls. However, surprisingly, TDE hosts lack any indication of *significant* recent merger activity from both morphological analysis and the machine learning merger classifier. Instead, our results reveal that TDE host galaxies in the green valley are approximately 1.5–3 times more likely to have bar-like or ringlike structures compared to their controls. Based on these results, we propose that bar-driven secular evolution, instead of mergers, likely dominates the recent evolution of the TDE hosts found in the green valley, which can simultaneously explain their distinctive nuclear properties and enhanced TDE rates.

## 1. Introduction

A tidal disruption event (TDE) occurs when a star ventures near a massive black hole (MBH) and is torn apart by tidal forces (M. J. Rees 1988). Approximately half of the stellar debris remains bound to the MBH, forming an accretion disk that produces a luminous flare through collision and subsequent accretion processes (C. R. Evans & C. S. Kochanek 1989; C. Bonnerot & N. C. Stone 2021; J. L. Dai et al. 2021; E. M. Rossi et al. 2021). TDEs offer a distinct opportunity to probe MBH demographics and gain insights into BH accretion physics (S. van Velzen et al. 2021). In recent years, the TDE detection rate has rapidly increased, due to advancements in instrumentation (S. Gezari 2021; E. Hammerstein et al. 2023). This surge has sparked growing interest in TDEs. A complete understanding of TDE astrophysics therefore requires characterizing both their host galaxy environments and the selection biases inherent to different observational surveys (K. D. French et al. 2016; J. Law-Smith et al. 2017). Upcoming surveys like Rubin/LSST will deliver a vast sample of TDEs, further emphasizing the need to advance our understanding of these events (K. Bricman & A. Gomboc 2018; S. Gezari et al. 2018).

Previous studies have revealed that TDE host galaxies are statistically distinct from the general galaxy population when matched on stellar mass and redshift. They exhibit: (1) higher central stellar concentrations (J. Law-Smith et al. 2017; O. Graur et al. 2018); (2) greater prevalence in the green valley (GV; J. Law-Smith et al. 2017; E. Hammerstein et al. 2021; Y. Yao et al. 2023); and (3) overrepresentation among E+A (poststarburst) galaxies, which are systems with weak H$\alpha$ emission (indicating suppressed star formation) and strong Balmer absorption (signaling a recent starburst; K. D. French et al. 2016; O. Graur et al. 2018).

The physical characteristics of TDE host galaxies likely have intrinsic connections to TDE rates. The occurrence rate of TDEs is fundamentally governed by the stellar dynamics in galactic nuclei, where stars interact through processes such as two-body relaxation (J. Wang & D. Merritt 2004). During these dynamical processes, the nuclear stellar structure—including the stellar population and density profiles—plays a critical role in setting the TDE rate (H. Pfister et al. 2020; N. C. Stone et al. 2020; J. N. Y. Chang et al. 2025). These works demonstrate that denser stellar environments in the vicinity of MBHs naturally enhance TDE rates.

Galaxy mergers have been suggested as compelling mechanisms for explaining some of these observed properties of TDE host galaxies (K. D. French et al. 2020b). Major mergers can distort galaxy shapes through tidal forces, drive radial gas inflows, and disrupt stellar rotation (D. Lynden-Bell 1967; A. Toomre 1977; R. De Propris et al. 2007; P. F. Hopkins et al. 2013; D. R. Patton et al. 2016). This triggers central starbursts and MBH accretion (L. Hernquist 1989; M. Sparre & V. Springel 2016; F. Renaud et al. 2022; S. Byrne-Mamahit et al. 2023, 2024; L. Ferreira et al. 2025) as well as stellar bulge growth (A. Brooks & C. Christensen 2016; S. L. Ellison & L. Ferreira 2026), thereby enhancing nuclear stellar densities. The energetic feedback following episodes of significantly enhanced central star formation and MBH accretion in major mergers is also a compelling pathway for making poststarburst galaxies (S. L. Ellison et al. 2022; S. Quai et al. 2023; S. Ellison et al. 2024). Interestingly, H. Pfister et al. (2019) incorporated basic TDE physics into a galaxy merger simulation and found that the TDE rates can indeed increase by orders of magnitude at certain phases after mergers.

It is expected that mergers should leave detectable signatures in galaxy structure, such as tidal tails or asymmetric features (A. Toomre & J. Toomre 1972; J. E. Barnes & L. Hernquist 1998; C. J. Conselice 2003; J. M. Lotz et al. 2008). Various studies have searched for observational evidence of mergers in the current sample of TDE host galaxies, but none have obtained conclusive evidence. For example, J. Law-Smith et al. (2017) found no significant enhancement in asymmetry among their sample of 10 TDE hosts using Sloan Digital Sky Survey (SDSS) imaging (K. N. Abazajian et al. 2009). However, this analysis was limited by the small TDE sample size, SDSS' relatively shallow depth, and poor spatial resolution, which could make faint merger signatures hard to detect (S. Wilkinson et al. 2024). K. D. French et al. (2020b) later examined four TDE hosts using higher-resolution Hubble Space Telescope (HST) imaging but similarly found no evidence for recent mergers.

While galaxy mergers remain a compelling scenario, other evolutionary pathways should also be considered to account for the unique characteristics of TDE host galaxies. Simulations and observations support the diversity of quenching pathways: T. A. Davis et al. (2019) find that poststarburst galaxies arise from a variety of mechanisms, including but not dominated by major mergers, while M. M. Pawlik et al. (2019) show that only a minority of poststarburst galaxies follow the classic "blue-to-red" merger track, with many experiencing star formation truncation without a dominant starburst. This opens the door for slower, less disruptive processes to be important mechanisms for creating galaxies with specific properties. In particular, secular evolution, which is the gradual internal transformation in relatively isolated galaxies, can also enhance galaxy nuclear stellar densities, despite being a much less violent process compared to galaxy mergers (J. Kormendy & R. C. Kennicutt 2004). Unlike mergers, secular evolution often preserves axisymmetric galaxy structures, providing an alternative route to the dense stellar cores favorable for TDEs.

In the study presented here, we conduct a detailed morphological analysis of TDE host galaxies to determine whether their observed structures are more consistent with recent merger activity or alternative mechanisms, such as secular evolution. For this study, higher-quality galaxy images are crucial, as previous work has demonstrated that merger features become undetectable as image quality deteriorates (C. Bottrell et al. 2019; R. McElroy et al. 2022; S. Wilkinson et al. 2024). We leverage the unprecedented sample size of SDSS (D. G. York et al. 2000) and the exceptional capabilities of the Dark Energy Camera Legacy Survey (DECaLS) and Ultraviolet Near-Infrared Optical Northern Survey (UNIONS; S. Gwyn et al. 2025). In particular, observations conducted using the Canada–France–Hawaii Telescope (CFHT) on Maunakea provide an excellent combination of depth, resolution, and wide-area coverage in the $r$ bands. These high-quality optical images have a strong record of revealing faint tidal features and disturbed morphologies (R. W. Bickley et al. 2021; S. Wilkinson et al. 2022; L. Ferreira et al. 2024), providing key signatures of recent galaxy interactions that may enhance TDE rates. To maximize the scientific return from UNIONS' superb data, we also combine these images with comprehensive structural measurements from S. Wilkinson et al. (2022) and merger classifications based on a hybrid machine learning approach (L. Ferreira et al. 2024, 2026).

The paper is structured as follows. We provide a description of the methodology in Section 2, where we introduce the imaging surveys we utilize, the details of the morphological parameters and classification method, our sample of TDE host galaxies, their control sample, the identifications of ring and bar structures, and the boundary of the star-forming main sequence (SFMS). We compare broad physical properties of TDE host galaxies to non-TDE hosts in Section 3.1 and contrast stellar concentration in Section 3.2 and merger indicators in Section 3.3 and Section 3.4. We then investigate the fractions of bars and rings in these galaxies in Section 3.5. We discuss the possible impact of bar-driven secular evolution on TDE host galaxies in Section 4 and summarize our results in Section 5.

## 2. Methods and Data

### 2.1. Imaging Data

We obtain images from SDSS, DECaLS, and UNIONS. SDSS represents one of the most comprehensive astronomical surveys, combining both imaging and spectroscopic observations, while DECaLS and UNIONS offer deeper and higher-resolution images.

#### 2.1.1. SDSS

SDSS is composed of imaging and spectroscopic surveys, carried out with a dedicated 2.5 m telescope at the Apache Point Observatory in Southern New Mexico. Its imaging camera contains 30 CCDs, with a median $r$-band seeing of $1\farcs32$ and a $5\sigma$ point source depth of 22.7 mag (D. G. York et al. 2000).

We utilize data from Data Release 7 (DR7; K. N. Abazajian et al. 2009), which includes data from three main surveys: the SDSS Legacy Survey, the Sloan Extension for Galactic Understanding and Exploration (SEGUE) survey, and the SDSS Supernova survey. The DR7 imaging data cover approximately 8423 deg$^2$ of the legacy sky and 3240 deg$^2$ of the SEGUE sky, while the spectroscopic data span 8200 deg$^2$ of the sky.

#### 2.1.2. DECaLS

DECaLS (R. D. Blum et al. 2016) is conducted using the Dark Energy Camera (DECam) located at the Cerro Tololo

Inter-American Observatory. It is mounted on the Victor M. Blanco 4m telescope. It contains 74 CCDs and images a 3 deg$^2$ field of view , with an $r$-band depth of 23.4 mag. It covers 14,000 deg$^2$ in the $g$, $r$, and $z$ bands. More importantly, it overlaps with SDSS in the region $-20° <$ decl. $< +30°$.

#### 2.1.3. UNIONS

UNIONS (S. Gwyn et al. 2025) is a collaboration of widefield imaging surveys of the northern hemisphere. UNIONS consists of the Canada–France Imaging Survey (CFIS), conducted at the 3.6 m CFHT on Maunakea; members of the Pan-STARRS team; and the Wide Imaging with Subaru Hyper Suprime-Cam of the Euclid Sky (WISHES) team. CFHT/CFIS is obtaining deep $u$- and $r$-band imaging; Pan-STARRS is obtaining deep $i$- and moderate–deep $z$-band imaging, and Subaru is obtaining deep $z$-band imaging through WISHES and $g$-band imaging through the Waterloo–Hawaii IfA $g$-band Survey (WHIGS). These independent efforts are directed, in part, to securing optical imaging to complement the Euclid space mission, although UNIONS is a separate collaboration aimed at maximizing the science return of these large and deep surveys of the northern skies.

UNIONS achieves exceptional image quality with median $r$-band seeing of $0\farcs6$, more than twice as sharp as SDSS, and reaches an impressive $5\sigma$ depth of 25.3 mag, making it more sensitive to faint sources and structures compared to SDSS. UNIONS covers approximately 5000 deg$^2$ in the $r$ band.

UNIONS provides significant advantages over SDSS for studying galaxy morphology and evolution. Its superior image quality and greater depth enable the detection of faint tidal features and low-surface-brightness structures that are often unresolved in SDSS data. These improvements allow for more precise morphological measurements, particularly in the area of galaxies where merger signatures and secular processes leave subtle imprints.

While UNIONS' footprint is limited to decl. $> 30°$, it is complemented by DECaLS, which provides similar quality imaging for decl. $< 30°$. Therefore, the combined high-resolution imaging from UNIONS and DECaLS, paired with SDSS' invaluable spectroscopic coverage, creates a powerful synergy that enhances our ability to investigate galaxy formation and dynamical evolution.

### 2.2. Measurements and Classification

In this section, we present the galaxy sample selection and derived properties for our TDE host analysis. As will be described in this section, we obtain a baseline sample of $n_{\rm SDSS} = 585{,}016$ galaxies with basic derived properties and measurements. We utilize complementary datasets including: (1) morphological measurements of both SDSS and UNIONS images; and (2) merger classification by the MUlti Model Merger Identifier (MUMMI; L. Ferreira et al. 2024, 2026) on UNIONS and DECaLS images. In addition, we also utilize UNIONS and DECaLS images for bar and ring classification. Figure 1 shows a schematic of the described data.

#### 2.2.1. Morphological Measurements

We start with the SDSS DR7 catalog (M. A. Strauss et al. 2002). We include the star formation rate (SFR) and velocity dispersion $\sigma$ from the MPA-JHU catalog (J. Brinchmann et al. 2004) and the total stellar mass $M_*$ from J. T. Mendel et al. (2014). We consider only SDSS galaxies with $z > 0.001$ and $M_* > 10^8 M_\odot$. This results in a sample with size $n_{\rm SDSS} = 585{,}016$.

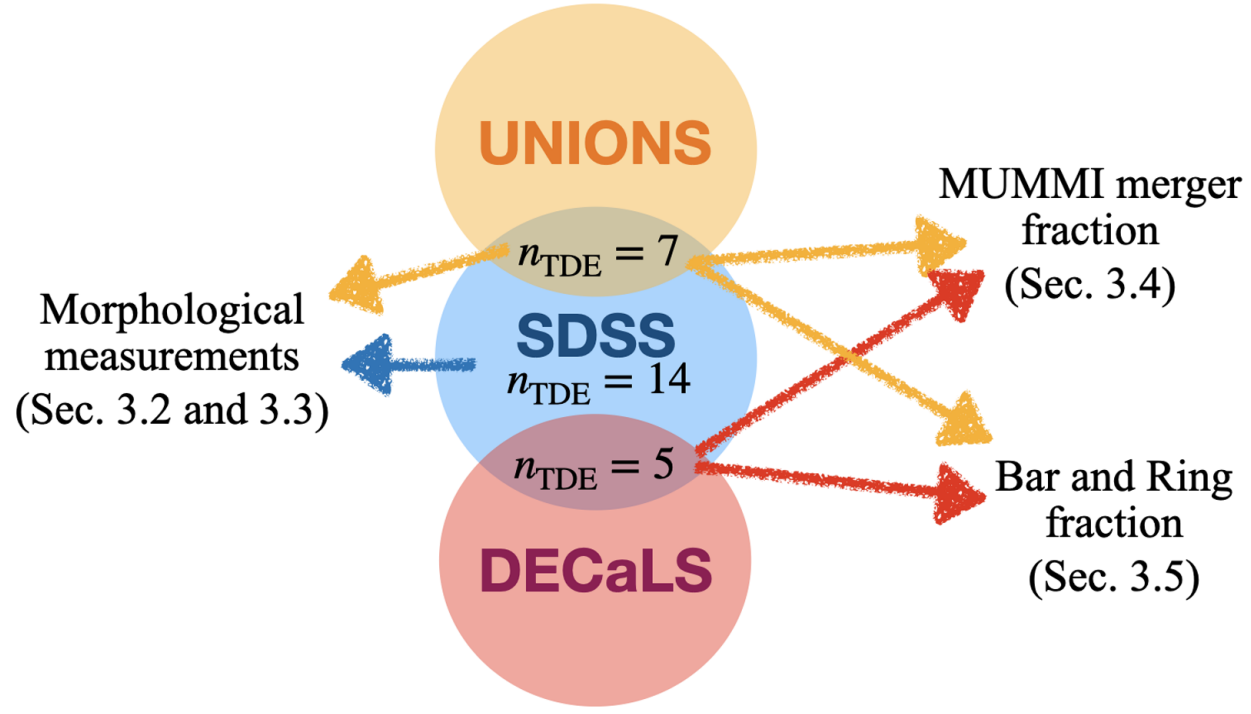

**Figure 1.** Schematic diagram illustrating the overlap of the image catalogs used for various analyses. The central blue region represents the initial SDSS spectroscopic catalog (with a sample of 14 TDE host galaxies). Overlaid are the UNIONS (orange) and DECaLS (red) galaxies, showing their respective overlaps: seven of the SDSS TDE hosts are also present in UNIONS, and five are also found in DECaLS. The color-coded arrows indicate which image catalog was utilized for each specific analysis discussed in Section 3.

We use morphology measurements from S. Wilkinson et al. (2022), computed with STATMORPH (V. Rodriguez-Gomez et al. 2019) and applied to both SDSS and UNIONS images. The concentration parameter $C_{\rm stat}$ is measured as follows:

$$C_{\rm stat} = 5\log\left(\frac{r_{80}}{r_{20}}\right), \quad (1)$$

where $r_{80}$ and $r_{20}$ are the radii containing 80% and 20% of the total flux, respectively. The asymmetry parameter $A_{\rm stat}$ is measured by rotating the image by 180°, subtracting the rotated image from the original, and then normalizing by the total flux. This total image asymmetry is then reduced by the background component:

$$A_{\rm stat} = \frac{\sum_{ij}|I_{ij} - I_{ij}^{180}|}{\sum_{ij}|I_{ij}|} - B, \quad (2)$$

where $I_{ij}$ and $I_{ij}^{180}$ are the flux of a pixel in the original orientation and after rotating by 180° about the center of the galaxy, respectively. Here, $B$ denotes the average asymmetry of the background. We note that there are differences in the background subtraction approaches between the surveys. To assess the impact of these differences, we conducted a comparative investigation and found no evidence of any systematic discrepancies between UNIONS and SDSS, aside from the improved depth offered by UNIONS. If we retain only valid STATMORPH measurements measured using SDSS images (by keeping data where both the asymmetry and concentration values are not "none"), this reduces the sample size to $n_{\rm stat} = 520{,}331$. Similarly, by keeping valid STATMORPH measurements measured with UNIONS images, the sample size reduced to $n_{\rm stat,UNIONS} = 198{,}626$.

We note that while there are other available morphological measurements, namely one measured by L. Simard et al. (2011), we use STATMORPH for the reason that it is applied homogeneously to both SDSS and UNIONS images.

### 2.2.2. Merger Classification

We complement traditional asymmetry-based merger identification methods with MUMMI (L. Ferreira et al. 2024, 2026). MUMMI employs a hybrid approach, combining convolutional neural networks and vision transformers, trained on realistic synthetic data from the illustrisTNG100-1 simulation (R. Weinberger et al. 2017; A. Pillepich et al. 2018). This framework enables the highly accurate classification of merger stages, achieving a purity rate of 95%. Crucially, due to its training data, MUMMI is designed to identify mergers primarily in galaxies with stellar mass $M_* > 10^{10} M_\odot$ and covering a mass ratio from 1:1 down to 1:10. Many of our TDE hosts are below $10^{10} M_\odot$ stellar mass. However, L. Ferreira et al. (2024) demonstrate that there is no brightness bias within the magnitude range in the SDSS galaxies used in MUMMI's classification (see their Figure 15). The merger fraction in the low-mass range (between $10^9$ and $10^{9.5} M_\odot$) is also consistent with the values reported in A. Guzmán-Ortega et al. (2023) for both simulated and observed galaxies. We also visually inspected MUMMI-classified images in this slightly "out-of-domain" stellar mass range to confirm that it accurately separates mergers and nonmergers. Although our TDE hosts are slightly below the targeted stellar mass range, these findings give us confidence that MUMMI is capable of classifying mergers within the stellar mass range relevant to our TDE host sample.

The catalog of mergers from UNIONS imaging is sourced from the published work by L. Ferreira et al. (2024, 2026). To complement UNIONS with imaging at decl. $< 30°$, we also utilize a separate catalog of mergers identified by MUMMI applied to DECaLS imaging. This MUMMI-DECaLS catalog is currently unpublished but will be presented in S. L. Ellison et al. (2026, in preparation).

In both the UNIONS and DECaLS cases, a galaxy's merger status is determined through a voting-based approach among different models. Specifically, we classify a galaxy as a merger if it receives the majority of the votes in the ensemble classification process. After crossmatching with the SDSS catalog, a total of $n_{\rm MUMMI,UNIONS}$ = 359,228 and $n_{\rm MUMMI,DECaLS}$ = 196,648 galaxies from UNIONS and DECaLS have majority merger votes from MUMMI, as well as SDSS spectroscopic measurements.

### 2.3. TDE Host Galaxy Sample

We compile our sample starting from the TDE catalogs in T. H. T. Wong et al. (2022), where the authors had already compiled a list of the most likely TDE candidates from T. Wevers et al. (2017, 2019), K. D. French et al. (2020b), S. van Velzen et al. (2021), and S. Gezari (2021). We complement the list with recent discoveries from E. Hammerstein et al. (2023). For each TDE, we crossmatched its coordinate to the nearest galaxy within a 3″ radius, using the SDSS DR7 spectroscopic catalog (M. A. Strauss et al. 2002) and the UNIONS catalog, ensuring reliable host galaxy associations for subsequent analysis. Our sample includes eight galaxies that overlap with the J. Law-Smith et al. (2017) catalog,[11] supplemented by six newly identified TDE host galaxies, totaling 14 TDE host galaxies. Their R.A. and decl. are presented in Table 1, and images are shown in Figure 2.

The mass of the BH ($M_{\rm BH}$) is a critical parameter in TDE physics, governing both the disruption radius of stars and the subsequent flare properties. We estimate $M_{\rm BH}$ through the $M_{\rm BH}$–$\sigma$ relation (J. Kormendy & L. C. Ho 2013),

$$M_{\rm BH} = 10^9 \times 0.309 \times (\sigma_e/200)^{4.38}, \quad (3)$$

prioritizing velocity dispersion ($\sigma$) measurements obtained from high-resolution Keck spectra (K. D. French et al. 2016) for 10 out of 14 of our TDE host galaxies, and falling back to values from the MPA-JHU catalog (J. Brinchmann et al. 2004) when unavailable. All $\sigma$ measurements from MPA-JHU were corrected to a standardized aperture using the relation from I. Jorgensen et al. (1995), accounting for variations in SDSS fiber sizes:

$$\sigma_e = \sigma_{v,MPA}/10^{-0.065\times\log(r_e/1.5)-0.013\times\log^2(r_e/1.5)}, \quad (4)$$

where $r_e$ is the effective radius from L. Simard et al. (2011). We acknowledge uncertainties in determining the BH mass, stemming from various sources of $\sigma$ and the intrinsic scatter of the scaling relation (±0.3 dex). Still, this hierarchical approach ensures consistent mass estimates across the TDE sample while prioritizing high-quality spectroscopic data when available.

### 2.4. Non-TDE Host Galaxy Control Sample

Key galaxy properties, such as stellar mass, correlate with $M_{\rm BH}$ and evolve with redshift (J. Kormendy & L. C. Ho 2013; S. Dattathri et al. 2025). A robust statistical comparison requires matching the TDE host sample to control galaxies with similar fundamental parameters. Therefore, we match our TDE host galaxies with galaxies of similar properties from SDSS. The following matching parameters are considered: redshift ($z$), BH mass ($M_{\rm BH}$), and total stellar mass of the host galaxy ($M_*$). We note that we do not match in SFR for our analysis of asymmetry and concentration, because SFR is strongly correlated with both of these measurements, at least for star-forming galaxies (H. M. Yesuf et al. 2021). Since mergers are also established drivers of enhanced asymmetries, central stellar densities, and SFR, controlling for SFR could inadvertently remove genuine differences related to merger activity, which is relevant to our analysis.

First, we define a specific range for each matching parameter (±0.1 dex for $M_{\rm BH}$ and $M_*$, and ±0.01 for $z$) and randomly select a sample of $N = 500$ galaxies within these ranges. If the number of galaxies within these ranges is insufficient, we then relax our criteria and incrementally expand the range: by ±0.05 dex for $M_{\rm BH}$ and $M_*$ (up to a maximum of ±0.25 dex), and by ±0.005 for $z$ (up to a maximum of ±0.025). This process continues until the sample size reaches $N = 500$.

Subsequently, for each TDE host galaxy with morphological parameter $X_{\rm TDE}$, we compare it with the corresponding measurement $X_{\rm control,j}$ from its control sample of 500 galaxies. Note that the control sample measurements were performed using the same survey as for the matched TDE host galaxy. To account for the variations in match quality, we weight each control measurement, following the method of D. R. Patton et al. (2016). Specifically, a control galaxy at $z_{\rm control,j}$ within a redshift tolerance of $z_{\rm tol}$ = 0.01 of its matched TDE host

[11] TDE #9 in J. Law-Smith et al. (2017) is classified as a QSO, and TDE #10 has no optical spectra listed under any source in SDSS. Hence, both are dropped from our sample.

**Table 1**
Properties of the 14 TDEs

| $i$ | TDE | R.A. (deg) | Decl. (deg) | $z$ | $\log(M_{BH}/M_{\odot})$ | Type | References |
|---|---|---|---|---|---|---|---|
| 1 | ASASSN-14ae* | 167.167150 | 34.097845 | 0.043671 | **5.96** | Optical/UV | K. D. French et al. (2020a) |
| 2 | ASASSN-14li | 192.063457 | 17.774013 | 0.020648 | **6.77** | X-ray | K. D. French et al. (2020a) |
| 3 | AT2018hyz (ASASSN-18zj) | 151.711958 | 1.692786 | 0.045798 | **6.20** | Optical/UV | K. D. French et al. (2020a) |
| 4 | AT2019azh (ASASSN-19dj) | 123.320625 | 22.648302 | 0.022299 | 6.58 | Optical/UV | E. Hammerstein et al. (2023) |
| 5 | AT2018dyk* | 233.283416 | 44.535690 | 0.036753 | **7.39** | Optical/UV | E. Hammerstein et al. (2023) |
| 6 | AT2020ocn* | 208.474173 | 53.997150 | 0.070509 | 6.76 | Optical/UV | E. Hammerstein et al. (2023) |
| 7 | AT2020wey | 136.357788 | 61.802544 | 0.027413 | 5.56 | Optical/UV | E. Hammerstein et al. (2023) |
| 8 | PTF09ge* | 224.263268 | 49.611373 | 0.064706 | **6.79** | Optical/UV | K. D. French et al. (2020a) |
| 9 | RBS1032 | 176.861244 | 49.716049 | 0.026087 | **5.81** | X-ray | K. D. French et al. (2020a) |
| 10 | SDSSJ0748* | 117.086119 | 47.203960 | 0.061533 | **7.61** | Optical/UV | K. D. French et al. (2020a) |
| 11 | SDSSJ1323* | 200.924890 | 48.450351 | 0.087540 | **6.62** | X-ray | K. D. French et al. (2020a) |
| 12 | SDSSJ1342 | 205.685065 | 5.515594 | 0.036644 | **6.55** | Optical/UV | K. D. French et al. (2020a) |
| 13 | SDSSJ1350* | 207.506230 | 29.269355 | 0.077726 | 7.92 | Optical/UV | K. D. French et al. (2020a) |
| 14 | iPTF15af | 132.117261 | 22.059302 | 0.078997 | **7.28** | Optical/UV | K. D. French et al. (2020a) |

**Note.** For each TDE, the table lists its index $i$, TDE name, R.A., decl., redshift $z$, BH mass $M_{BH}$, TDE type, and reference. The seven TDE host galaxies that are also found in UNIONS are highlighted with "*." $M_{BH}$ values derived using $\sigma$ from Keck are in bold; all other $M_{BH}$ values are derived using $\sigma$ from the MPA-JHU catalog.

galaxy at $z_{TDE}$ will have a redshift weight of

$$w_{z,j} = 1 - \frac{|z_{TDE} - z_{control,j}|}{z_{tol}}. \quad (5)$$

The overall weight is the product of the weights of all matching parameters:

$$w_j = w_{z,j} \cdot w_{M_{BH},j} \cdot w_{M_*,j}, \quad (6)$$

where $w_{M_{BH},j}$ and $w_{M_*,j}$ are the weights corresponding to the matching parameters $M_{BH}$ and $M_*$. With these weights, we compute the weighted mean of the controls:

$$\bar{X}_{control} = \frac{\sum_j w_j \cdot X_{control,j}}{\sum_j w_j}. \quad (7)$$

We then compute the difference between the TDE host galaxy's parameters and the weighted mean of its controls as

$$\Delta X = X_{TDE} - \bar{X}_{control}. \quad (8)$$

We discuss how the uncertainties are obtained in Appendix B.

### 2.5. Identification of Ring and Bar Structures

Figure 3 shows two examples of the TDE host galaxies in which we are able to obtain both SDSS (top) and DECaLS or UNIONS (bottom) images. This comparison demonstrates the improved image quality that DECaLS and UNIONS provide (resolution and depth). They reveal a wealth of morphological detail about TDE host galaxies that was previously inaccessible in SDSS data. Taking PTF09ge as an example, its ring structure is particularly striking in the UNIONS image compared to that of SDSS. It displays two bright nuclear patches surrounded by a faint outer ring, a configuration reminiscent of the $R_1$ or $R_1'$ morphology (characterized by a 180° winding of spiral arms that form a ring or pseudo-ring) in R. J. Buta (2017). This structure aligns with HST/WFC3 observations by K. D. French et al. (2020a), which confirmed both the blue outer ring and potential bar features of this particular galaxy. Another example is ASASSN-19dj, where faint ringlike or shell-like features are barely discernible in the SDSS image but appear significantly more prominent in DECaLS imagery.

We developed a systematic approach to classify the presence of bars and rings in our galaxy sample. Our analysis began with 12 TDE host galaxies with high-resolution imaging (seven identified in UNIONS and five in DECaLS). For each TDE host galaxy, J.C. compiled images of 10 randomly selected control galaxies from the same survey, drawn as established in Section 2.4. We performed the comparison both with and without additional matching in SFR; however, we chose to use control galaxies matched in SFR during visual classification, because this reduces systematics. This ensures we compare galaxies with similar star formation activity, better isolating effects related to the TDE host. In contrast, not controlling for SFR could conflate differences stemming from the intrinsic properties of GV versus SFMS populations, or those of passive clouds, complicating attributions of observed differences to TDE activity at that evolutionary stage. N.C. subsequently randomly shuffled these images and provided them to C.B., R.K.C., and J.D. for independent, double-blind visual classification, ensuring that the classifiers were unaware of which galaxies were TDE hosts. The vote assignments followed a three-tiered confidence system: confident presence of the feature, uncertain presence of the feature, or no detected feature. For the TDE host galaxies, we quantified the presence of features by assigning a score of 1 for confidently identified features and a score 0.5 for uncertain features, effectively weighting uncertain classifications accordingly. A TDE host galaxy was considered to possess a ring/bar if its averaged score was above 0.5. For the control galaxies, we determined the fraction with confidently identified features as a lower limit, and included uncertain classification as an upper limit, thus providing a range for the estimated feature prevalence.

We also obtained ring and bar fractions from the citizen projects GALAXY CRUISE and Galaxy Zoo DESI. GALAXY CRUISE (M. Tanaka et al. 2023) is a citizen science project that contains morphological classification, which is based on the second data release from the Hyper Suprime-Cam, Subaru Strategic Program (HSC-SSP) of 20,686 galaxies at $z < 0.2$. We crossmatch this GALAXY CRUISE catalog with our SDSS catalog, finding 14,990 matches within 3″. Unfortunately, our list of TDE host galaxies did not overlap with this crossmatched sample, meaning no GALAXY CRUISE classifications are available for them. We are interested in the third question in the survey, where one can pick one out of four options ("Ring," "Fan," "Tail," or "Distorted") for the features of the galaxy. Following the classification criteria from R. Shimakawa et al. (2024), we consider the galaxy to be ringed if

$$(\mathrm{P(interact)} > 0.5) \bigcap (\mathrm{P(ring)} = \max(\mathrm{P}_i)) \tag{9}$$

and nonringed if

$$(\mathrm{P(spiral)} < 0.5) \bigcap (\mathrm{P(ring)} < \max(\mathrm{P}_i)). \tag{10}$$

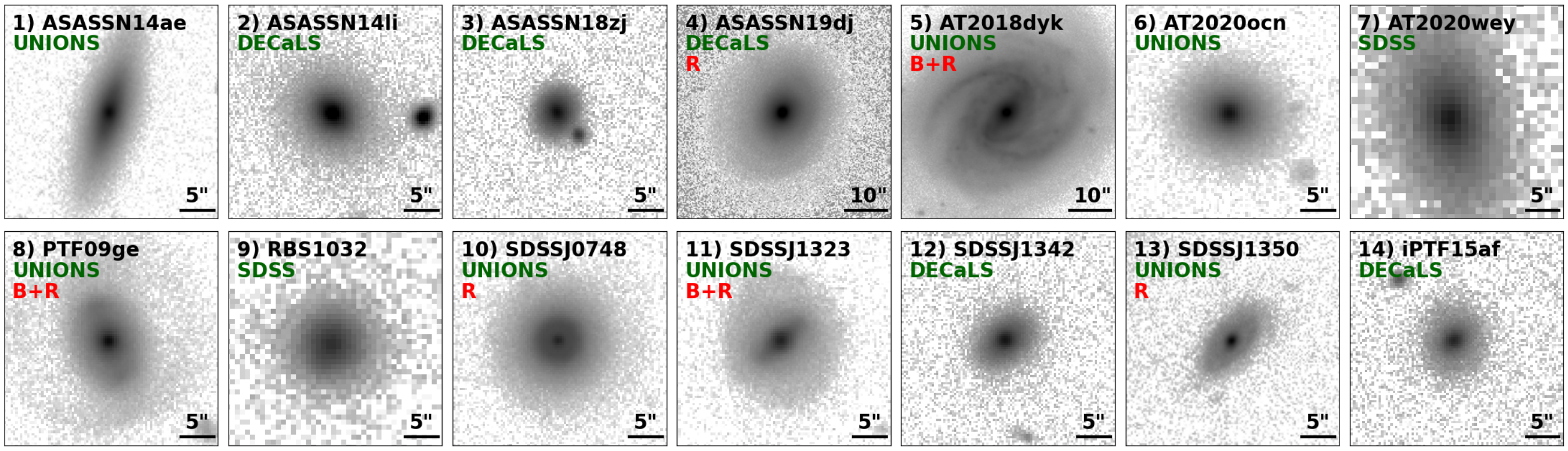

**Figure 2.** Representative $r$-band images of the 14 TDE host galaxies in Table 1. Images are sourced hierarchically: from UNIONS, if available; otherwise from DECaLS; and finally from SDSS. Each panel indicates the TDE index, TDE name, its originating survey, and the structural classification ("R" for ring and "B" for bar) in the top-left corner, with an angular size reference provided in the bottom-right corner.

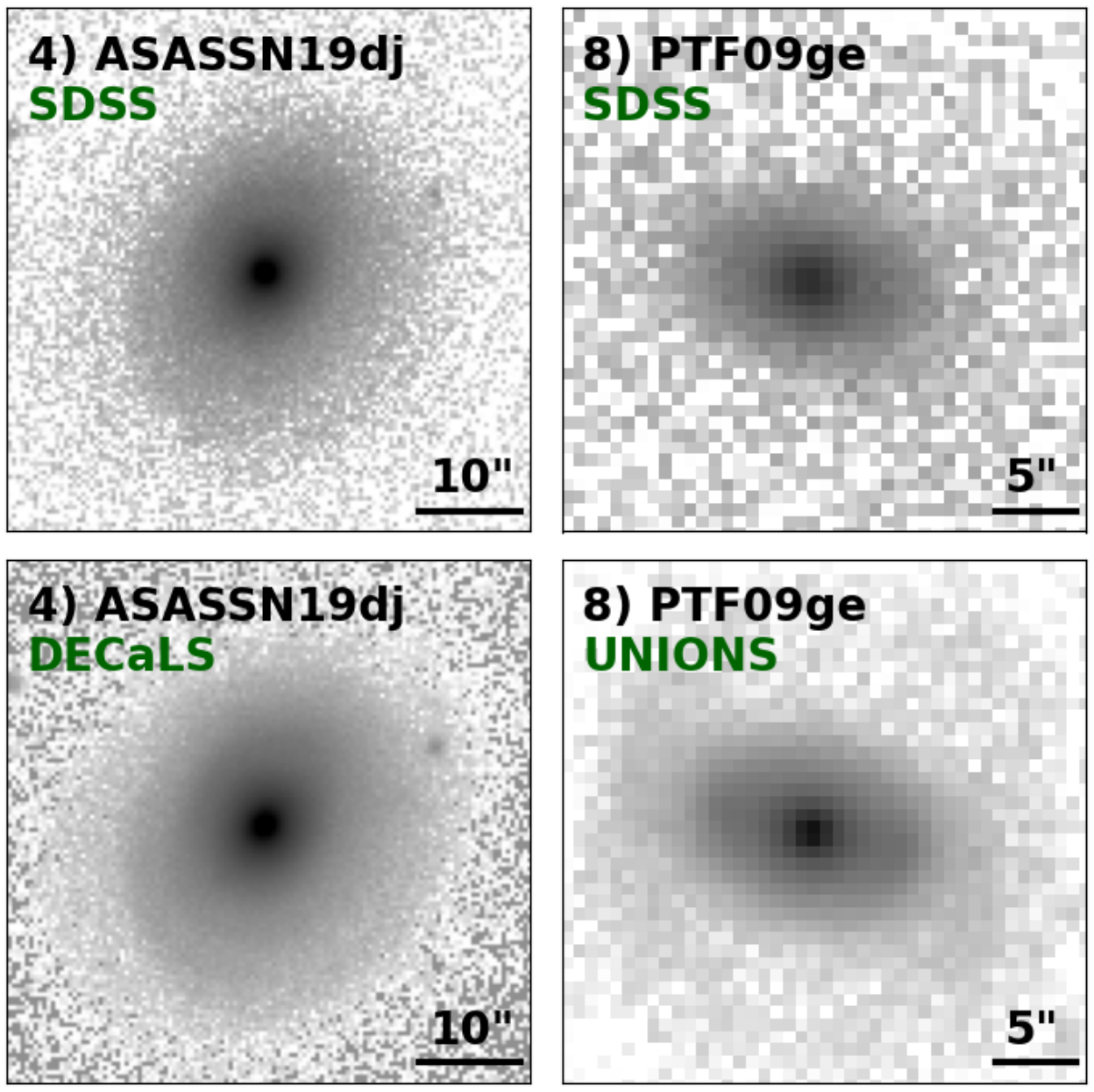

**Figure 3.** Comparison of $r$-band images for TDE host galaxies ASASSN-19dj (left) and PTF09ge (right) from different surveys. The top row display images from SDSS DR7, while the bottom row shows images from DECaLS (for ASASSN-19dj, bottom left) and UNIONS (for PTF09ge, bottom right). The side-by-side comparison clearly demonstrates the significantly enhanced resolution of the DECaLS and UNIONS imaging relative to SDSS. An angular size reference is provided in the bottom-right corner of each panel.

Here, P(*spiral*) and P(*interact*) give the probability of the galaxy showing spiral structures and evidence of interaction, based on Questions 1 and 2, respectively. $P_i \in \{\mathrm{P}(ring), P(fan), P(tail), P(distorted)\}$ represents the respective probabilities assigned in Question 3 that the galaxy should exhibit each type of specific morphological disturbance. These individual probabilities satisfy the condition $\sum_i P_i = 1$.

On the other hand, we compute the bar fraction using the morphology classification from M. Walmsley et al. (2023). These classifications are made using a deep learning model trained on Galaxy Zoo, containing votes for the DESI Legacy Survey (A. Dey et al. 2019) DR8 and historical Galaxy Zoo DECaLS votes. We crossmatch this result with the SDSS catalog and find 110,069 matches with a $3''$ matching tolerance, none of which overlap with our list of TDE host galaxies. Regarding bar features, each galaxy has a predicted fraction of votes in three classes: strong bar, weak bar, or no bar. We label each galaxy with the class with the highest vote fraction, considering the fraction of galaxies with strong bars to be a lower limit and the combined fraction of galaxies with either strong or weak bars to be the upper limit of the predicted bar fraction.

### 2.6. SFMS and GV

The observed incidence of TDEs in the GV (J. Law-Smith et al. 2017; E. Hammerstein et al. 2021; Y. Yao et al. 2023) necessitates a precise operational definition to distinguish these transitional systems from both star-forming and quiescent populations. We base our GV definition on the SFMS, the well-established correlation between $M_*$ and SFR for star-forming galaxies (J. Brinchmann et al. 2004; K. G. Noeske et al. 2007). Following M. Donnari et al. (2019), we implement an iterative fitting procedure. We first bin galaxies into 0.2 dex intervals of $M_*$. For each mass bin, we iteratively calculate a clipped mean SFR and standard deviation $\sigma$, then linearly extrapolate the SFR–$M_*$ relation until the SFR is affected by the quiescent region (where adjacent bins show $>1\sigma$ deviation in the mean SFR, typically cut at $\log M_* = 10.2$ ). After excluding outliers ($>2\sigma$ below the trend), we repeat this process until achieving convergence (slope variation $<0.001$ between iterations). The SFMS can be described by

$$\log(\mathrm{SFR}) = 0.72 \log(M_*) - 7.03, \tag{11}$$

and the lower SFMS boundary is established at $1\sigma$ below this converged relation.

Within this framework, we define star-forming galaxies as those lying above the lower SFMS boundary, while the GV occupies the crucial transitional space starting from the lower SFMS boundary to 1 dex below it. This operational definition captures the GV's evolutionary significance as the intermediate phase between blue, star-forming galaxies and red, quiescent systems (S. Salim 2014), while remaining consistent with established classification schemes (J. Law-Smith et al. 2017; S. A. Dodd et al. 2021). The resulting GV boundaries provide the necessary foundation for our subsequent analysis of TDE environments and can be described by the following equation:

$$0.83\log(M_*) - 8.67 < \log(\mathrm{SFR}) < 0.83\log(M_*) - 7.67. \tag{12}$$

We refer to Section 3.5 for a visual of the boundaries.

## 3. Results

### 3.1. General Properties of TDE Hosts

We compare our sample of 14 SDSS TDE host galaxies (Table 1) with the general SDSS galaxy population in Figure 4. The gray contours show the general galaxy distribution, while our TDE hosts are marked as red circles with their corresponding indices from Table 1. Our TDE hosts have BH masses $10^6$–$10^7 M_\odot$, consistent with T. Wevers et al. (2017), and relatively low stellar masses, following the expected $M_{\mathrm{BH}}$–$M_*$ relation. Additionally, all TDEs in our sample have low redshifts ($z = 0.02$–$0.09$), reflecting current survey detection limits. Their SFRs place them mostly in the GV between the SFMS and quiescent galaxies. Specifically, nine of the TDE hosts are located in the GV, three reside on the SFMS, and the remaining two are quiescent. Their bluer bulge colors ($B_g - B_r < 1$) indicate younger nuclear stellar populations. The TDE hosts display intermediate Sérsic indices ($n \sim 2$–$4$), suggesting pseudo-bulges or disturbed morphologies, and span a wide range of bulge-to-total (B/T) ratios. These trends (lower $M_{\mathrm{BH}}$ and $M_*$, bluer bulges, intermediate bulge structures, and GV SFRs) align closely with the findings of J. Law-Smith et al. (2017), reinforcing that TDEs preferentially occur in galaxies with recent central star formation and distinct nuclear properties.

### 3.2. Concentration of Stellar Density

The central stellar concentration has been shown to significantly relate to the rate of TDEs in galaxies (K. D. French et al. 2020b; H. Pfister et al. 2020; J. N. Y. Chang et al. 2025).

The top panel of Figure 5 displays the STATMORPH concentration parameters $C_{\mathrm{stat}}$ of the TDE host galaxies (stars) and their control samples (crosses) in relation to $M_{\mathrm{BH}}$, $M_*$, and $z$. The control sample measurements were weighted-averaged over 500 samples and obtained using the same survey as their matched TDE host, color-coded with yellow for UNIONS and blue for SDSS. Overall, $C_{\mathrm{stat}}$ increases steadily with $M_{\mathrm{BH}}$ and $M_*$, aligning with the expectation that the concentration increases with total mass, reflecting the transition from disk-dominated to bulge-dominated systems in the mass sequence (X.-F. Deng 2013).

The bottom panel of Figure 5 shows $\Delta C_{\mathrm{stat}} = C_{\mathrm{stat,TDE}} - \bar{C}_{\mathrm{stat,control}}$ for each TDE. TDE hosts with UNIONS images are represented in blue, and others with SDSS images are represented in orange. Additionally, a histogram of the $\Delta C_{\mathrm{stat}}$ distribution is provided on the right, with the shaded region indicating the average of $\Delta C_{\mathrm{stat}}$ with a $1\sigma$ uncertainty. Details of the uncertainties are discussed in Appendix B.

One can see that $\Delta C_{\mathrm{stat}}$ is consistently above 0. The mean value across all 14 TDE hosts is $\Delta C_{\mathrm{stat}} = 0.32 \pm 0.03$, indicating that TDE hosts are $\sim$16% more concentrated than the control galaxy samples. When focusing solely on the UNIONS data, $\Delta C_{\mathrm{stat}}$ slightly increases to $0.37 \pm 0.05$. This significant enhancement in $\Delta C$ remains robust even when controlling for additional morphological parameters, such as different metrics of size ($r_{50}$, $r_{80}$), during the control sample selection. Interestingly, high-$\Delta C_{\mathrm{stat}}$ TDE hosts are predominantly found in intermediate-mass systems, with no obvious trend observed with $z$. It is difficult to determine whether this observation is skewed by a few systems with particularly high $C_{\mathrm{stat}}$—namely, ASASSN-19dj and SDSSJ1350. Further data are needed to fully understand these trends.

### 3.3. Asymmetry of Galaxy Structure

The structure of a galaxy, especially its asymmetry, can provide important insight into its recent and ongoing merger activity (C. J. Conselice 2003; D. R. Patton et al. 2016). In this section, we investigate the asymmetry of our TDE host galaxies compared to their control samples, to assess the role that mergers may play in their evolution.

The top panel of Figure 6 plots the STATMORPH asymmetry parameter ($A_{\mathrm{stat}}$) for the TDE host galaxies (star symbols) and their control samples (cross symbols) against $M_{\mathrm{BH}}$, $M_*$, and $z$. As in Figure 5, the measurements are color-coded by survey: blue for UNIONS and orange for SDSS. We emphasize that TDE hosts are compared to controls from the same survey, ensuring a robust comparison. The bottom panel presents the difference in asymmetry, $\Delta A_{\mathrm{stat}} = A_{\mathrm{stat,TDE}} - \bar{A}$stat, control. The green shaded region denotes the mean $\Delta A$stat across all TDE hosts, while the histogram on the right shows the corresponding distribution of the residuals.

Surprisingly, the results indicate no significant evidence of increased asymmetry in our TDE host galaxies. The averaged $\Delta A_{\mathrm{stat}}$ across the sample is $-0.02 \pm 0.01$, suggesting that, on average, TDE hosts do not exhibit more asymmetric features than their control counterparts. While a high central concentration might, in principle, obscure outer asymmetries, we confirm that our finding of no significant asymmetry enhancement in TDE hosts remains robust even when the control samples are additionally matched on parameters such as concentration and $r_{80}$ (see Appendix C). Focusing on the UNIONS sample yields a lower $\Delta A_{\mathrm{stat}}$ of $-0.06 \pm 0.02$. This reduction demonstrates how improved image quality accentuates the lack of asymmetric structures among TDE hosts relative to the control sample.

Our results appear to suggest a negative correlation between $\Delta A_{\mathrm{stat}}$ and $M_*$. However, closer examination reveals that this apparent relationship is primarily attributable to differences in image quality, as evidenced by the clear separation of $\Delta A_{\mathrm{stat}}$ estimates for SDSS and UNIONS hosts. Improving the image quality generally reveals more faint asymmetric structures in galaxies, but preferentially in galaxies that are intrinsically asymmetric (C. Bottrell et al. 2019; M. D. Thorp et al. 2021; S. Wilkinson et al. 2024). Indeed, a comparison of the UNIONS and SDSS asymmetry measurements shows that, on average, the UNIONS asymmetries are about 0.05 higher than the SDSS asymmetries.

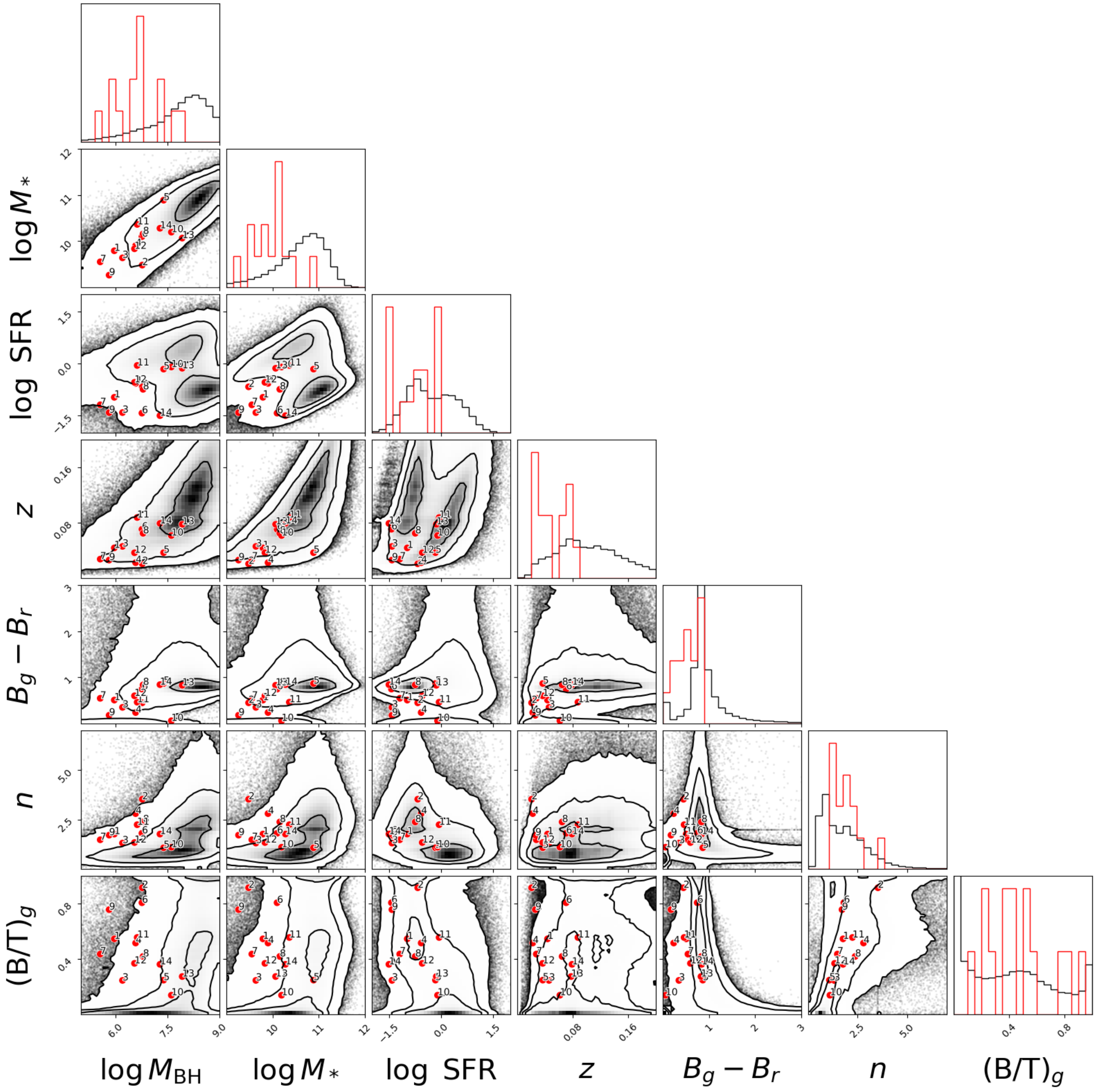

**Figure 4.** Corner plots comparing parameter distributions of the general SDSS galaxy population (contours) and our TDE host galaxy sample (red circles). Each red circle marker is annotated with its corresponding index from Table 1. The parameters displayed (from left to right) are: BH mass $M_{\rm BH}$, stellar mass $M_*$, SFR, redshift $z$, bulge color $B_g - B_r$, Sérsic index $n$, and B/T ratio $(B/T)_g$. The contours represent the $1\sigma$, $2\sigma$, and $3\sigma$ confidence levels of the SDSS galaxy distribution. These plots reveal that TDE hosts tend to have lower BH and stellar masses, reside at low redshifts, exhibit bluer bulge colors indicative of younger stellar populations, and possess intermediate structural properties such as Sérsic index and bulge dominance.

The upper panels of Figure 6 show that the UNIONS controls are more asymmetric than the SDSS controls, while the TDE host asymmetries are less distinct between SDSS and improved UNIONS imaging. These preferentially boosted control asymmetries in UNIONS result in suppressed average $\Delta A_{\rm stat}$ for UNIONS TDE hosts relative to SDSS. TDE hosts with UNIONS imaging coincidentally have higher stellar masses and are found at higher redshifts, resulting in an apparent trend. If we had UNIONS images for all TDE hosts, we could potentially eliminate this apparent correlation.

We perform a similar analysis using asymmetry measurements from L. Simard et al. (2011) in Appendix C. TDE host galaxies are not enhanced in this asymmetry measurement either. To robustly characterize the morphological features of TDE hosts, future studies should incorporate images from combined surveys and develop a systematic approach to measuring these morphological measurements across surveys (e.g., E. Sazonova et al. 2024).

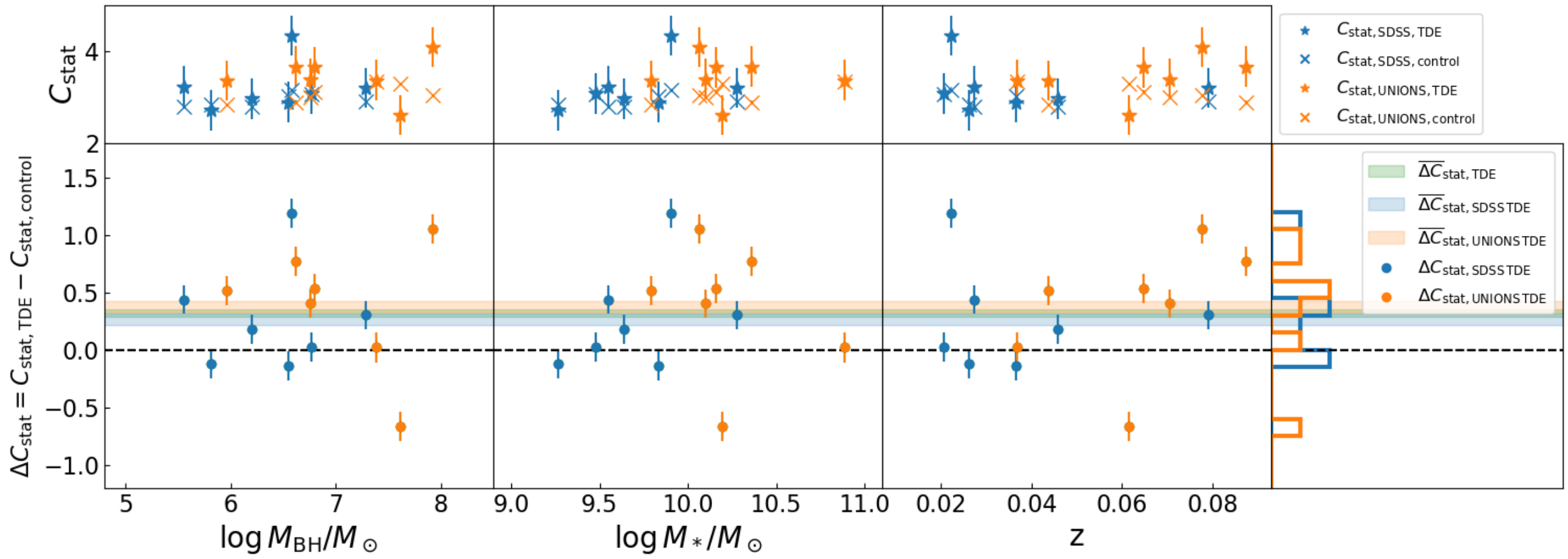

**Figure 5.** Concentration parameter analysis for TDE host galaxies and their control samples. The top panel presents the STATMORPH concentration parameter $C_{stat}$ for TDE host galaxies (star symbols) and their control sample (cross symbols) as a function of $M_{BH}$, $M_*$, and $z$. The measurements are based on UNIONS images (orange) when available, and SDSS images (blue) otherwise. Importantly, the control sample measurements were weighted-averaged over 500 samples and consistently derived from the same survey as their TDE host galaxy. The bottom panel displays the difference in concentration, $\Delta C_{stat} = C_{stat,TDE} - \bar{C}_{stat,control}$, with $1\sigma$ error bars. The shaded region indicates the mean $\Delta C_{stat}$ for the UNIONS (orange), SDSS (blue), and combined (green) subsets. A dashed line marks $\Delta C_{stat} = 0$. The rightmost panel provides a histogram of $\Delta C$ for these different samples. Overall, these results demonstrate that TDE host galaxies exhibit a significantly higher concentration parameter compared to their control samples, with an average $\Delta C_{stat} = 0.32 \pm 0.03$, implying TDE host galaxies are $\sim$16% more concentrated.

### 3.4. Postmerger Classification

Another way to assess whether these TDE host galaxies have likely undergone recent galaxy mergers is to use the UNIONS MUMMI (L. Ferreira et al. 2024, 2026). In the MUMMI catalog, each galaxy is classified using 20 independently trained models to determine whether it is a merger. The galaxy's merger status is then assigned based on the number of models that vote it as a merger. For our analysis, a galaxy is classified as a merger if more than 10 out of the 20 models vote it as such. This criterion is applied consistently to both the TDE host galaxies and the control samples.

Among the 12 TDE host galaxies that are covered by MUMMI, all were classified as nonmergers. In contrast, 4.9% of the UNIONS controls and 5.2% of the DECaLS controls were identified as mergers. When combining both datasets, the overall merger fraction among the controls is approximately 5.03%, whereas none of the TDE hosts are classified as mergers.

Given the small sample size and the low expected merger rate ($\sim$5% from the controls), we acknowledge the statistical uncertainty inherent in these estimates. To quantify this, we employ a Bayesian approach, with a prior merger rate of 5% modeled using a beta distribution $\beta(\alpha_{prior} = 1, \beta_{prior} = 19)$. For the TDE hosts, the number of "successes" (0 mergers out of 12 galaxies) can be modeled with a binomial likelihood. Combining the prior with the likelihood gives a posterior of $\beta(\alpha_{prior}, \beta_{prior} + 12)$. This approach indicates that the true merger rate is around 3.1%, with a 95% highest density interval of 0.1%–9.2%. While the limited sample size prevents us from excluding modest increases in merger activity, we can rule out a twofold increase in merger activity at the $1\sigma$ confidence level. Notably, given the high prevalence of poststarburst features in TDE hosts (K. D. French et al. 2016), and given that 60%–75% of poststarburst galaxies are mergers or postmergers (S. L. Ellison et al. 2022; S. Ellison et al. 2024), we would have expected significantly higher merger activity than our results indicate.

The results from both morphology analysis and MUMMI classification provide no strong evidence for a *significant* enhancement in merger activities among TDE host galaxies relative to their controls. The heightened concentration in TDE hosts without a matching increase in asymmetry raises questions about the underlying processes. The absence of expected asymmetrical features suggests recent merger-driven nuclear bursts of star formation are unlikely. Additionally, the MUMMI merger classifications do not indicate a clear excess of ongoing interactions in the TDE host population. Therefore, alternative mechanisms should be explored to understand how galaxies enhance their nuclear stellar concentration.

### 3.5. Bar and Ring Fractions

In this section, we investigate the bar and ring structures in the TDE host galaxies and compare them with their control galaxies. Following the method mentioned in Section 2.5, we identify three TDE hosts with bar-like structures and six TDE hosts with ringlike structures among the 12 TDE host galaxies that have quality imaging. Among the seven TDE hosts in the GV, three display bar-like structures and four display ringlike structures. Given the small sample size, we carefully estimated confidence intervals following E. Cameron (2011), to ensure their reliabilities. We show the location of the TDE hosts in the SFR versus $M_*$ plot along with their classification in Figure 7. As for the control samples, 13.6%–20.0% of the galaxies display bar-like structures (the lower limit set by confidently identified bar structures, and the upper limit set by including uncertain bar identifications). At the same time, 16.9%–24.0% of the controls displayed ringlike structures. When limited to the controls of GV TDEs, this barred fraction shifted slightly to 11.9%–18.0%, and the ringed fraction shifted to 15.2%–23.0%.

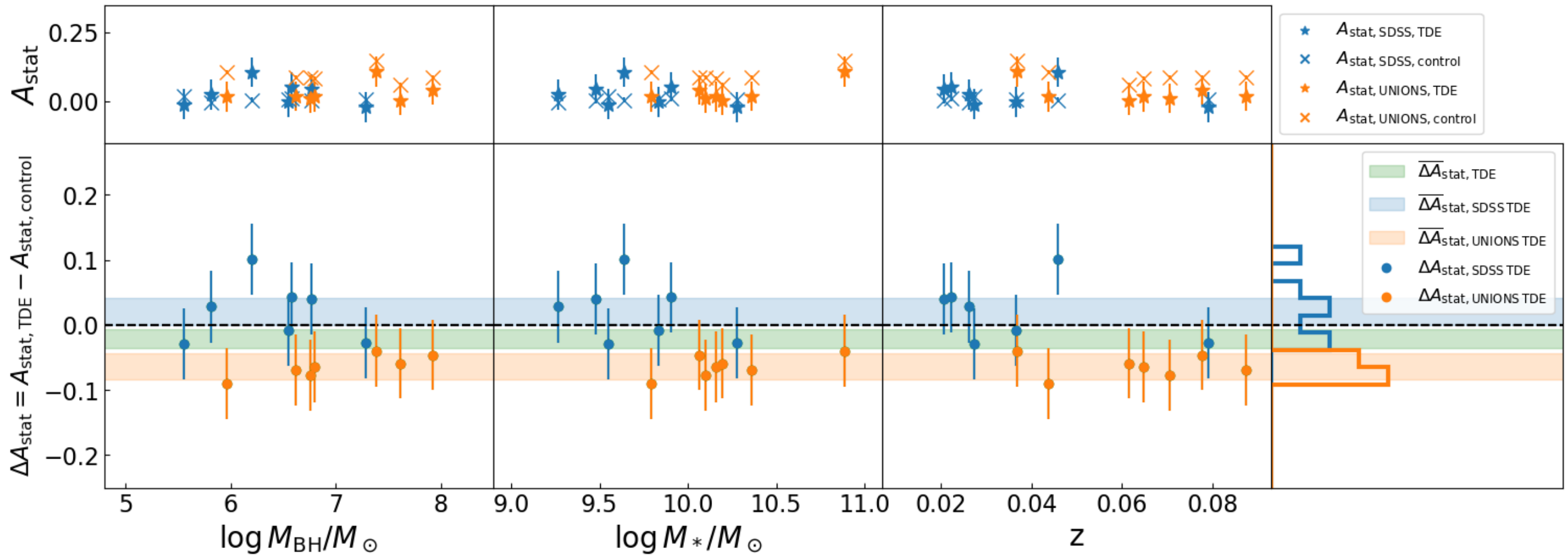

**Figure 6.** Asymmetry parameter analysis for TDE host galaxies and their control samples. The top panel presents the STATMORPH asymmetry parameter $A_{\rm stat}$ for TDE host galaxies (star symbols) and their control samples (cross symbols) as a function of $M_{\rm BH}$, $M_*$, and $z$. As in Figure 5, measurements are based on UNIONS images (orange symbols) when available, and SDSS images (blue symbols) otherwise. The control sample measurements were weighted-averaged over 500 samples and consistently derived from the same survey as their TDE host galaxy. The bottom panel displays the difference in the asymmetry parameter $\Delta A_{\rm stat} = A_{\rm stat,TDE} - A_{\rm stat,control}$, with $1\sigma$ error bars. The shaded regions indicate the mean $\Delta A_{\rm stat}$ for the UNIONS (orange), SDSS (blue), and combined (green) subsets. A dashed line marks $\Delta A_{\rm stat} = 0$. The rightmost panel provides a histogram of the $\Delta A_{\rm stat}$ distribution for these different samples. Overall, these results show that TDE host galaxies, especially the higher-resolution UNIONS sample, display no indication of enhanced asymmetry (and actually slightly reduced asymmetry) compared to their control sample.

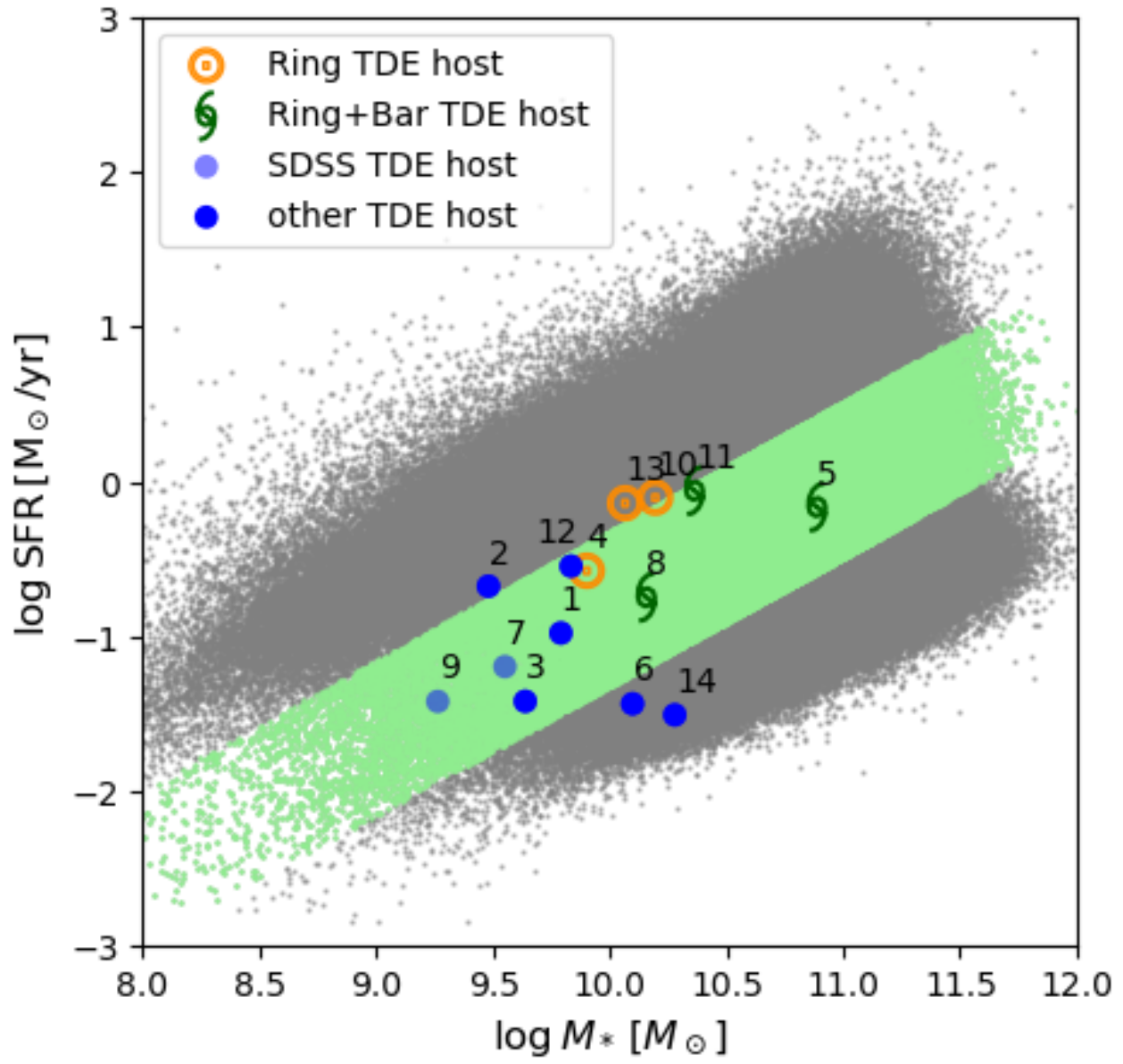

**Figure 7.** Locations of 14 TDE host galaxies in the SFR versus $M_*$ plane. The different symbols represent the morphological classifications (see Section 2.5) of the TDE host galaxies. The opacity of the symbols indicates the image resolution used for classification: more opaque symbols correspond to hosts with DECaLS or UNIONS images, while more transparent symbols denote hosts for which only SDSS images were available. The GV region, as defined in Equation (12), is shaded in green, with other regions shown in gray. Notably, among the TDE host galaxies within the GV and observed with high-resolution imaging, three out of seven exhibit bar structures, and four out of seven display ring structures.

We compute reference ring and bar fractions from the citizen projects GALAXY CRUISE and Galaxy Zoo DESI, respectively. These fractions are calculated from control samples matched in $M_{\rm BH}$, $M_*$, and $z$ to the TDE hosts, as described in Section 2.4. For controls matched to the full list of 14 TDE hosts in our sample, the ring fraction ranges between 10%–35%, while for controls matched specifically to GV TDE hosts, this ranges between 11%–39%. Similarly, the bar fraction ranges between 8%–34% for all controls, increasing to 10%–39% for the GV TDE host controls.

The ring and bar fractions derived from citizen projects exhibit broader ranges than those from our own visual inspection. Our internal classifications lie predominantly in the lower parts of these ranges, suggesting that our criteria are somewhat more conservative. However, because the classification process was conducted in a double-blind manner, any systematic differences in classification strictness should affect both samples similarly. We therefore expect that the relative difference in the ring and bar fractions between TDE hosts and controls remains robust.

Figure 8 summarizes our findings on bar and ring fractions, revealing a significant enhancement of these structures in TDE host galaxies compared to their control counterparts. Specifically, the TDE host galaxy sample exhibits a bar fraction of 25.0%, which is elevated by a factor of 1.5 relative to the control sample (average bar fraction of 16.8%), though still within statistical agreement. The enhancement in the ring structures is more pronounced (we make an additional conservative estimate of the enhancement by comparing the ratio between the $1\sigma$ lower limit of the TDEs and the controls' upper limit; E. Cameron 2011). TDE host galaxies show a ring fraction of 50%, which corresponds to an enhancement by a factor of 2.4 relative to the controls (conservatively, by a factor of 1.5).

These differences are even more striking within the GV. In this transitional region, the TDE host galaxy bar fraction rises to 42.9%, indicating a 2.9 times (1.6 times) enhancement relative to the GV control galaxies. Similarly, the TDE ring fraction in the GV is 57%, showing a factor of 3 (1.7) enhancement compared to the GV control sample. Overall, these statistics underscore a strong correlation between these internal morphological features and TDE activity, particularly for galaxies within the GV. Our results demonstrate that TDE host galaxies are significantly more likely to harbor bar and ring structures than typical galaxies of similar stellar mass and color.

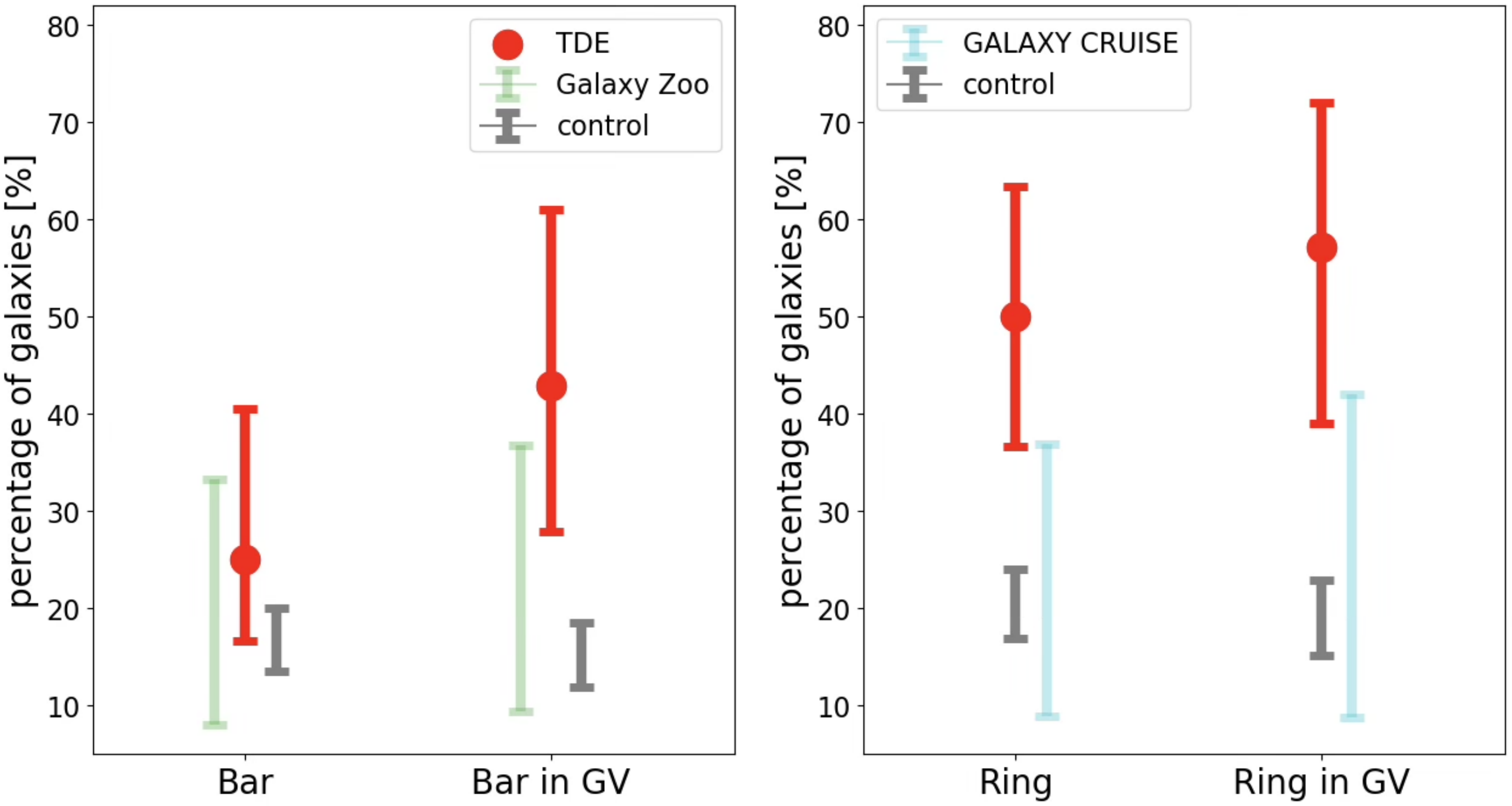

**Figure 8.** Bar (left panel) and ring (right panel) fractions for TDE host galaxies and various control samples. The bar/ring fractions for our TDE host galaxy sample, determined through visual classification, are indicated by the red bars. The gray bars represent the corresponding fractions for their matched control samples, classified using an identical visual method. These are shown for both the overall galaxy population and specifically for those within the GV. For comparison, the bar and ring fractions from the Galaxy Zoo (light green) and GALAXY CRUISE (light blue) control samples are also displayed. Overall, these results indicate that TDE host galaxies are more likely to possess bar or ring structures than their control samples, with this enhancement being more pronounced among TDE hosts located in the GV.

## 4. Discussion

While our TDE host galaxies exhibit no enhanced merger activity compared to their controls, we find that a significant proportion of the TDE host galaxies in our sample display barlike and ringlike structures. This suggests that secular processes could be an alternative driver for the connection between the poststarburst status and enhanced central stellar concentrations of TDE host galaxies.

It is important to acknowledge potential observational biases that may influence these results. Many of the nonbarred or seemingly featureless galaxies in our sample could, in fact, host undetected structural features. Recent high-resolution studies with the James Webb Space Telescope (JWST) have revealed that certain morphological structures can be missed or unresolved in lower-resolution surveys (V. Kuhn et al. 2024; Z. A. Le Conte et al. 2026a, 2026b). Moreover, our own findings show that some TDE hosts classified as compact or featureless in SDSS images exhibit prominent bars or rings when observed with deeper, higher-resolution data from UNIONS or DECaLS. This suggests that current survey sensitivities impose a lower limit on our ability to detect such features, and the true fraction of barred and ringed galaxies may be underestimated. While advanced techniques such as generative forward modeling can be employed to recover morphological structures that are otherwise obscured or unresolved in lower-quality data (e.g., A. Adam et al. 2025), such methods were beyond the scope of our current analysis. Instead, we endeavored to mitigate these observational biases by carefully constructing unbiased control samples for our TDE hosts, matching them on $z$, $M_{\rm BH}$, $M_*$, and, critically, imaging quality.

Nevertheless, for the TDE hosts in which bar or ring features are present, secular evolution driven by bars provides a compelling physical framework. Secular evolution, characterized by gradual internal changes in relatively isolated galaxies, plays a critical role in shaping galactic nuclei. Bars, in particular, generate nonaxisymmetric gravitational potentials that torque gas toward the center (E. Athanassoula 1992; B. G. Piner et al. 1995; P. Englmaier & O. Gerhard 1997), producing radial shocks and resonant interactions that drive material inward (W. W. Roberts et al. 1979; P. Verwilghen et al. 2024). Critically, resonance interactions between the bar and gas not only drive central inflows, but also concentrate material into rings (R. J. Buta 2017; M. C. Sormani et al. 2018). Recent work by M. Frosst et al. (2025) further supports this picture, demonstrating that barred galaxies host more massive central BHs than unbarred controls, likely due to the efficiency of the bar-driven gas inflow in fueling both star formation and BH growth.

As bars funnel gas inward, they can trigger nuclear starbursts (C. H. Heller & I. Shlosman 1994; W. W. Roberts et al. 1979; J. Kormendy & R. C. Kennicutt 2004; S. L. Ellison et al. 2011; R. J. Buta 2017). Whether this inflow also efficiently feeds active galactic nuclei (AGNs) remains an open question. While some studies find a correlation between bars and AGN activities (L. C. Ho 2005; S. Alonso et al. 2018), others report no significant connection (M. A. Galloway et al. 2015; A. D. Goulding et al. 2017).

The stellar feedback from the triggered nuclear starburst (e.g., supernova, winds) may subsequently deplete and/or heat gas reservoirs (C. C. Hayward & P. F. Hopkins 2017; H. He et al. 2023), while AGN feedback may further reduce the availability of star-forming gas (J. R. Mullaney et al. 2015; T. T. Shimizu et al. 2015). Notably, the structure of the bars also influences star formation in complex ways: shear forces in dust lanes can suppress massive star formation, while shocks enhance it (A. Zurita et al. 2004). Collectively, these mechanisms can leave poststarburst signatures in galaxies.

Thus, secular evolution driven by large-scale bars provides a compelling framework for explaining several observed properties of TDE host galaxies. Bars can funnel gas inward, to enhance central stellar densities and trigger nuclear starbursts (and potentially accelerate AGN activity), resulting in poststarburst galaxies. Crucially, this process increases the central stellar concentration without inducing significant global asymmetry, aligning with our findings. Furthermore, bars can dynamically enhance TDE rates: stars trapped in bar-aligned orbits can develop high eccentricities (E. Athanassoula 1992), and subsequent perturbations could efficiently scatter these stars into the loss cone, where they will eventually be disrupted (D. Merritt & M. Y. Poon 2004).

However, bars alone may not be the sole mechanism responsible for the formation of a dense nuclear stellar distribution and subsequent TDE events. Bars are correlated with stellar mass and galaxy morphology, being more prominent in massive late-type spiral galaxies, while weaker bars are observed in low-mass early-type dwarfs and irregular galaxies (P. Erwin 2018). Despite this correlation, TDEs have been observed in low-mass galaxies, necessitating further investigation to enhance our current understanding.

## 5. Summary

In this study, we have examined the morphology and physical properties of TDE host galaxies based on imaging data from SDSS, DECaLS, and UNIONS. From a reference catalog of known TDE events, we identified 14 matching host galaxies from the SDSS spectroscopic sample. We compared the host properties to matched non-TDE-hosting controls in redshift, stellar mass, and BH mass. We compared the morphologies of the TDE host galaxies to their controls using visual classifications, structural measurements, and a machine learning–based merger classifier. We provide a summary of our findings below.

1. TDE host galaxies exhibit a notable increase in concentration compared to their control sample ($\Delta C_{\rm stat} = 0.32 \pm 0.03$), indicating they are ~16% more concentrated. This suggests an enhancement in central stellar density near the MBH.
2. Our analysis of TDE host galaxies reveals a lack of enhancement in asymmetry in comparison to their control sample, with $\Delta A_{\rm stat} = -0.02 \pm 0.01$. Additionally, there are no indications that these TDE hosts are recent merger remnants or currently undergoing interactions, based on merger classifications from the machine learning model MUMMI. This surprising result does not favor the merger-driven origin of TDE hosts.
3. TDE host galaxies show enhanced bar-like and ringlike structures. Notably, among GV galaxies, the fractions of bars and rings are enhanced by a factor of 1.5–3 compared to their controls. This suggests that bar-driven secular evolution may play a key role in funneling gas toward the galaxy center, thereby increasing the central stellar density and potentially triggering TDEs.

TDE rates are governed primarily by the innermost region of the host galaxy. How this region connects to the larger-scale structure remains an important and open question. The claim that TDEs preferentially inhabit postmerger galaxies is largely built on the statistical links between TDE hosts and poststarburst galaxies and between poststarburst galaxies and mergers. Crucially, however, poststarburst galaxies are not merger-only phenomena. They can also result from other channels (Z. Cheng et al. 2024). Furthermore, recent studies have revealed that TDE hosts frequently exhibit extended emission-line regions (EELRs; T. Wevers & K. D. French 2024; M. Pursiainen et al. 2026). These EELRs are more directly associated with fading AGNs (K. D. French et al. 2023; T. Wevers & K. D. French 2024). Although AGN activity can be triggered by mergers, this is not required. Secular processes such as bar-driven gas inflow can also ignite nuclear activity. Indeed, the origin of these EELRs and their relationships to both nuclear stellar dynamics and large-scale galactic morphology remain to be better studied (A. Mummery et al. 2025). Compounding this, deep HST imaging studies have searched for direct morphological evidence of ongoing or recent mergers in a very small sample of TDE hosts but found no clear signatures (K. D. French et al. 2020b), which is consistent with our results. Taken together, these studies highlight uncertainties in the prevailing view that most observed TDEs inhabit postmerger environments. This leads us to consider an alternative scenario: a distinct internal mechanism for TDE triggering driven by the secular evolution of the stellar disk. Such evolution can modify the stellar dynamics in the nuclear region, enhancing TDE rates while simultaneously producing ring and bar features that may serve as observational signatures of this channel.

Establishing robust, statistically significant links between TDE incidence and specific large-scale galactic features, however, will require significantly larger samples of TDEs with well-characterized host galaxy properties. At present, our understanding remains limited by sample size: only roughly 10–20 observed TDEs currently have imaging suitable for the morphological analysis presented here. Expanding this analysis to larger, uniformly characterized datasets is therefore essential, both for confirming the emerging trends and for deepening our physical understanding of the processes that govern TDE rates. Future deep, high-resolution, and high-cadence imaging from next-generation facilities will be critical in this endeavor. The Vera C. Rubin Observatory (Ž. Ivezić et al. 2019; F. B. Bianco et al. 2022), for instance, is expected to identify approximately 1000 TDEs per year (S. van Velzen et al. 2019; K. Bricman & A. Gomboc 2020). Moreover, Rubin's unprecedented depth will extend TDE host galaxy samples to higher redshifts, where galaxy mergers are more frequent. Host galaxy analysis of the substantial TDE samples provided by next-generation surveys will be crucial for disentangling the relative contributions of internal (secular) and external (merger-driven) processes to TDE rates across cosmic time.

## Acknowledgments

We thank I. Arcavi, K. Auchettl, S. Dodd, I. Ebrova, N. Jiang, J. Lim, I. Mendel, E. Ramirez-Ruiz, T. Wevers, and Q. Wu for useful comments and discussions. We thank the participants of the TDE FORUM (Full-process Orbital to Radiative Unified Modeling) online seminar series for their inspiring discussions. We are also grateful to N. Chang and T. Kwan for their help in the visual classification process. J.C., L.D., and R.K.C. acknowledge support from the National Natural Science Foundation of China and the Hong Kong Research Grants Council (NSFC/RGC JRS N_HKU782/23 and RGC GRF 17314822). R.K.C. would like to acknowledge the financial support provided by the Anusandhan National Research Foundation (ANRF), a statutory body of the Department of Science and Technology (DST), Government of India, through the National Post-Doctoral Fellowship (NPDF) [Grant No. PDF/2025/004682]. C.B. gratefully acknowledges support from the Forrest Research Foundation. R.Y. acknowledges support by the National Natural Science Foundation of China (grant Nos. 12425302 and 12373008), by the Research Grant Council of Hong Kong (GRF 14303123), by the Jockey Club Charities Trust through the JC STEM Lab of Astronomical Instrumentation, and by the CUHK Direct Grant.

We are honored and grateful for the opportunity to observe the Universe from Maunakea and Haleakala, which both have cultural, historical, and natural significance in Hawaii. This work is based on data obtained as part of the Canada–France Imaging Survey, a CFHT large program of the National Research Council of Canada and the French Centre National de la Recherche Scientifique. Based on observations obtained with MegaPrime/MegaCam, a joint project of CFHT and CEA Saclay, at the Canada–France–Hawaii Telescope (CFHT), which is operated by the National Research Council (NRC) of Canada, the Institut National des Science de l'Univers (INSU) of the Centre National de la Recherche Scientifique (CNRS) of France, and the University of Hawaii. This research used the facilities of the Canadian Astronomy Data Centre operated by the National Research Council of Canada with the support of the Canadian Space Agency. This research is based in part on data collected at the Subaru Telescope, which is operated by the National Astronomical Observatory of Japan. Pan-STARRS is a project of the Institute for Astronomy of the University of Hawaii, and is supported by the NASA SSO Near Earth Observation Program under grants 80NSS-C18K0971, NNX14AM74G, NNX12AR65G, NNX13AQ47G, NNX08AR22G, and 80NSSC21K1572 and by the State of Hawaii.

## Appendix A
## The 14 SDSS TDEs' Morphological Parameters

In this section, we show the morphological parameters of our TDE host galaxies in Table A1.

**Table A1**
Morphological Parameters of the 14 SDSS TDEs

| $i$ | TDE | $C_{\mathrm{stat,UNIONS}}$ | $C_{\mathrm{stat,SDSS}}$ | $C$ | $A_{\mathrm{stat,UNIONS}}$ | $A_{\mathrm{stat,SDSS}}$ | $R_A$ | $V_{\mathrm{MUMMI}}$ |
|---|---|---|---|---|---|---|---|---|
| 1 | ASASSN-14ae | 3.35263 | 3.08042 | 0.496 | 0.017033 | 0.007253 | 0.016 | 0 |
| 2 | ASASSN-14li | ⋯ | 3.07758 | 0.747 | ⋯ | 0.044505 | 0.009 | 2 |
| 3 | AT2018hyz | ⋯ | 2.95760 | 0.433 | ⋯ | 0.106074 | 0.006 | 0 |
| 4 | AT2019azh | ⋯ | 4.33293 | 0.526 | ⋯ | 0.053525 | 0.030 | 0 |
| 5 | AT2018dyk | 3.35835 | 3.22714 | 0.296 | 0.108391 | 0.035370 | 0.042 | 0 |
| 6 | AT2020ocn | 3.38016 | 2.89894 | 0.411 | 0.010670 | −0.014597 | 0.010 | 0 |
| 7 | AT2020wey | ⋯ | 3.22147 | 0.470 | ⋯ | −0.010219 | 0.014 | – |
| 8 | PTF09ge | 3.64364 | 3.16977 | 0.413 | 0.017687 | −0.018834 | 0.020 | 0 |
| 9 | RBS1032 | ⋯ | 2.70987 | 0.476 | ⋯ | 0.025407 | 0.017 | – |
| 10 | SDSSJ0748 | 2.60593 | 2.73312 | 0.409 | 0.003300 | 0.006259 | 0.015 | 0 |
| 11 | SDSSJ1323 | 3.65326 | 3.29657 | 0.509 | 0.019084 | −0.063164 | 0.015 | 0 |
| 12 | SDSSJ1342 | ⋯ | 2.88064 | 0.445 | ⋯ | −0.001452 | 0.023 | 0 |
| 13 | SDSSJ1350 | 4.07896 | 4.20559 | 0.385 | 0.042012 | −0.007431 | 0.013 | 0 |
| 14 | iPTF15af | ⋯ | 3.19621 | 0.353 | ⋯ | −0.020717 | 0.012 | 0 |

**Note.** The parameters (from left to right) are: index $i$, TDE name, STATMORPH concentration measured using the UNIONS image $C_{\mathrm{stat,UNIONS}}$, STATMORPH concentration measured using the SDSS image $C_{\mathrm{stat,SDSS}}$, the concentration measurement from L. Simard et al. (2011) $C$, STATMORPH asymmetry measured using the UNIONS image $A_{\mathrm{stat,UNIONS}}$, STATMORPH asymmetry measured using the SDSS image $A_{\mathrm{stat,SDSS}}$, the residual asymmetry $R_A$ from L. Simard et al. (2011), and MUMMI's number of votes (out of 20 models) for merger identification $V_{\mathrm{MUMMI}}$.

## Appendix B
## Error Propagation

In this section, we detail the error calculations used in Figures 5 and 6. Unfortunately, the morphological measurement does not technically come with uncertainties. We estimate the uncertainty of an individual measurement $\sigma_m$ based on E. Sazonova et al. (2024), taking the rms error of the asymmetry measurements from pairs of observations of the same galaxy from images of similar depth as our typical uncertainty.

The weighted mean is

$$\bar{X}_{\rm control} = \frac{\sum_j w_j \cdot X_{\rm control,j}}{\sum_j w_j}, \quad \text{(B1)}$$

with a weighted variance

$$\sigma_w^2 = \frac{\sum_j w_j \cdot (X_{\rm control,j} - \bar{X}_{\rm control})^2}{\sum_j w_j}, \quad \text{(B2)}$$

and thus the standard error of the mean is

$$\sigma_{\rm SEM}^2 = \frac{\sum_j w_j^2 \sigma_w^2}{\left(\sum_j w_j\right)^2}. \quad \text{(B3)}$$

The difference between the TDE host and its controls is

$$\Delta X = X_{\rm TDE} - \bar{X}_{\rm control}. \quad \text{(B4)}$$

Hence, it has an uncertainty of

$$\sigma_{\Delta X}^2 = \sigma_m^2 + \sigma_{SEM}^2. \quad \text{(B5)}$$

## Appendix C
## Other Measures of Asymmetry and Concentration

We include additional asymmetry and concentration analysis in this appendix. Figure C1 shows the asymmetry analysis, as a function of $M_{\rm Bh}$, $M_*$, and $z$, when including the STATMORPH concentration $C_{\rm stat}$ as an additional matching parameter. The resulting averaged $\Delta A_{\rm stat} = -0.00 \pm 0.01$ is consistent with our findings from the asymmetry analysis in Section 3.3.

We also include the analysis for the concentration $C$ (c1_g) and residue asymmetry $R_A$ (ra1_3_g) from L. Simard et al. (2011) as a function of $M_{\rm Bh}$, $M_*$, and $z$ in Figure C2.

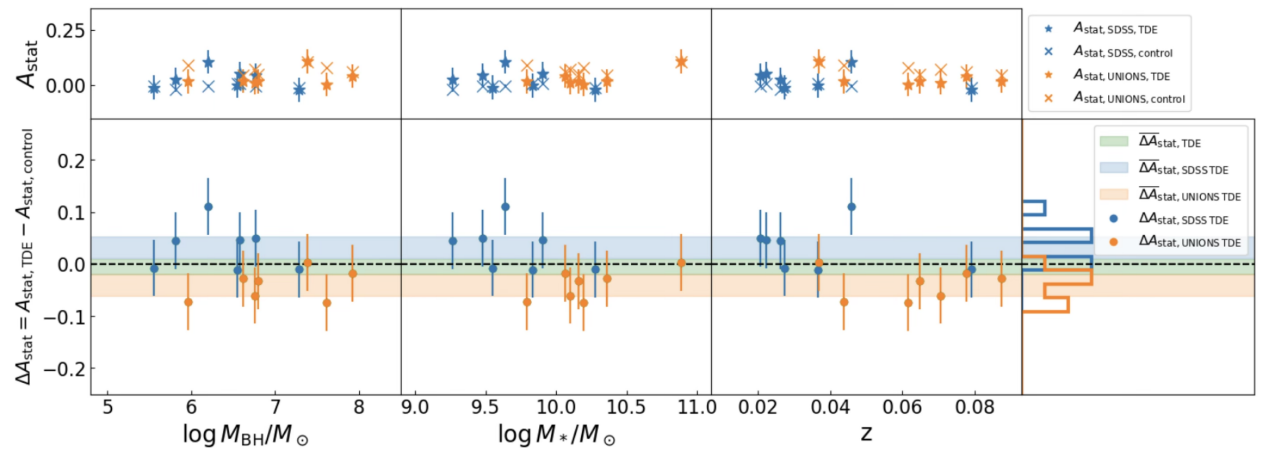

**Figure C1.** The same analysis in Section 3.3 on the asymmetry $A_{\rm stat}$, with the concentration $C_{\rm stat}$ included as an additional matching parameter for the controls. The symbols remain in the same style as in Figure 5. The result ($\Delta A = -0.00 \pm 0.01$) is consistent with our findings from the asymmetry analysis in Section 3.3.

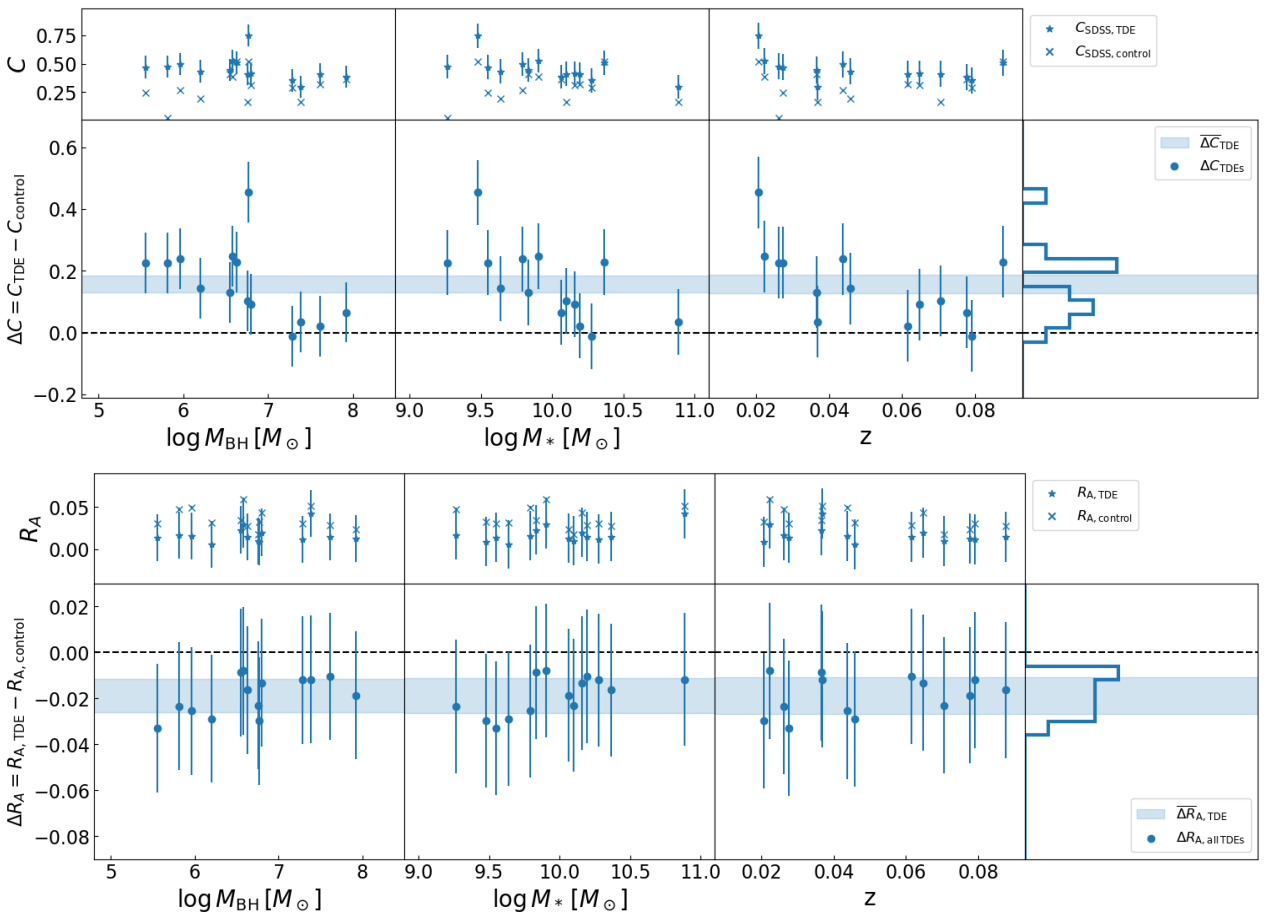

**Figure C2.** The same analysis in Section 3.2 on the concentration $C$ (top) and residual asymmetry $R_A$ (bottom) from L. Simard et al. (2011). The symbols remain in the same style as in Figure 5. These measurements are done using SDSS images only, hence they are only in blue. Asymmetries are not enhanced (slightly suppressed) and concentrations are enhanced in TDE hosts relative to controls. These outcomes agree qualitatively with our results using STATMORPH measurements.

## ORCID iDs

Janet N.Y. Chang https://orcid.org/0009-0004-2575-1924
Connor Bottrell https://orcid.org/0000-0003-4758-4501
Lixin Dai https://orcid.org/0000-0002-9589-5235
Rudrani Kar Chowdhury https://orcid.org/0000-0003-2694-933X
Meng Gu https://orcid.org/0000-0002-4267-9344
Renbin Yan https://orcid.org/0000-0003-1025-1711
Leonardo Ferreira https://orcid.org/0000-0002-8919-079X
Sara L. Ellison https://orcid.org/0000-0002-1768-1899
Scott Wilkinson https://orcid.org/0000-0002-3303-4077
Thomas de Boer https://orcid.org/0000-0001-5486-2747